\documentclass[12pt]{article}
\usepackage[dvips]{color}
\usepackage{epsfig}
\usepackage{amsmath}
\usepackage{cite}
\usepackage{color}
\usepackage{subfigure}

\begin{document}
\begin{center}
\Large{\bf Logarithmic corrected $f(R)$ gravitational model and swampland conjectures}\\
\small \vspace{1cm} {\bf J. Sadeghi$^{\star}$\footnote {Email:~~~pouriya@ipm.ir}}, \quad
{\bf E. Naghd Mezerji$^{\star}$\footnote {Email:~~~e.n.mezerji@stu.umz.ac.ir}}, \quad
{\bf S. Noori Gashti$^{\star}$\footnote {Email:~~~saeed.noorigashti@stu.umz.ac.ir}}, \quad
\\
\vspace{0.5cm}$^{\star}${Department of Physics, Faculty of Basic
Sciences,\\
University of Mazandaran
P. O. Box 47416-95447, Babolsar, Iran}\\
\small \vspace{1cm}
\end{center}
\begin{abstract}
In this paper, we use corrected $f(R)$ gravitational model which is polynomial function with a logarithmic term. In that case, we employ the slow-roll condition and obtain the number of cosmological parameter. This help us to verify the swampland conjectures which is guarantee the validation of low - energy quantum field theory. The obtained results shown that the corresponding model is consistent with the  swampland conjectures. Also the upper and lower limit of the parameter \textcolor{red}{$n$} are \textcolor{blue}{0.15} and \textcolor{blue}{0.0033}. Finally, by using scalar spectrum index $ n_{s} $ and tensor to scalar ratio $ r $ relations and compared with Planck 2018 empirical data, we obtain  the coefficients $\alpha$,$\beta$ and $\gamma$. Also, the corresponding results are creaked by several figures, literature and also plank 2018 data.
\end{abstract}
\section{Introduction}

As we know weak gravity conjecture (WGC) is a strong conjecture on gravitational coupling theories which shows that gravity is the weakest force in any theory consistent with quantum gravity\cite{Palti,Arkani,Medved}. There are some evidence for weak gravity conjecture which including holography, cosmic censorship,  entropy, unitary and causality. In order to study swampland and landscape, for the simplicity we have distinguish between general quantum gravity with string theory\cite{Brennan,Lin,Kallosh,Urbano}. If we want to understand the abstract concept of the swampland, we have to understand how to distinguish between  low energy effective  field theories  that are in the landscape from those in the swampland. The most important things here is WGC, because it provides a very powerful tool for distinguishing between landscape and swampland\cite{Urbano,Palti,Kallosh,Brennan}.  We note here a set of consistent low energy effective  field theories, which is compatible to string theory is called a landscape\cite{Murayama,Garg,Gar,Vafa}. We mention that also, at the low energies landscape is surrounded by  wider space called swampland, which is  consistent with quantum gravity. So, for more reviews about swampland, one can see Ref.s \cite{Urbano,Palti,Murayama}.
But, at high energies the whole space is landscape, which is consistent with quantum gravity. Generally, one can say that there are  different conditions for the verifying  that a low energy theory is  swampland or landscape\cite{Das,Linder}. One of the conditions that has been studied in recent years for the specifying about swampland is \cite{Huang,Matsui}
 \begin{equation}\label{1}
 \frac{|\Delta\phi|}{M_{pl}}\leq \Delta \sim1
 \end{equation}
and
 \begin{equation}\label{2}
 M_{pl}\frac{|V'(\phi)|}{V}\geq c \sim 1
 \end{equation}
 where $M_{pl}$ is Planck mass, we assume approximately equal  one, $c$ and $\Delta$ are positive constants as  order of unity \cite{Kinney,Garg,Schimmrigk,Cheong}. $V(\phi)$ is also the potential which is coming normalized canonical field $\phi$.  The criteria of swampland have been examined in various gravitational theories,  which can be found in the Ref.s.\cite{Yi,Brahma,Artymowski}.

The first condition of above mentioned equation actually satisfies the single field inflation,  but the second condition creates some problems for the single field system. Recently, new swampland conjecture models have been proposed\cite{Garg, Ooguri,Dutta},  this new models introduce a scalar field potential associated with a self-consistent $UV$ compleat which  must be  satisfied by  following two conditions,
 \begin{equation}\label{3}
8(M_{pl}\frac{|V'|}{V})\geq c
 \end{equation}\\
and
\begin{equation}\label{4}
(M_{pl}^{2}\frac{|V''|}{V})\leq -c'
 \end{equation}

where $c$ and $c'$ are  unit order and constants .
As you know, one of the dark energy models based on modified gravity is actually called $f(R)$ gravity .  $f$ is a function of scalar Ricci, and in general we have $ f=F+R $\cite{Capozziello,Nojiri,Carroll}. Such modified gravity will be suitable candidate for  describing   dark energy  and cosmic acceleration.
So, in this paper we examine the applications of WGC  constraint on a deformed Starobinsky gravity\cite{Starobinsky,Codello}. The purpose of this paper, we employ one part of WGC model as  swampland condition. In that case,  we take advantage from $f (R)$ modified gravity  and examine the inflation theory. Here one can say that,  the obtained results from swampland condition can be  compared  by  experimental data.  Also,  we note here that several researchers worked with some simple form of $f(R)$, which are given by \cite{Arapoglu,Orellana,Sadeghi,Channuie}. But, we take another $f(R)$ which is including both polynomial  and a logarithmic form. In that case, we  apply the corresponding WGC condition and verify  some suitable parameters in $f(R)$ gravity. All above information about the swampland condition on the cosmological model give us motivation to organized  the paper as follows. In  section 2, we introduce the function$ f (R)$ and consider our corresponding model. In section 3, by using the above-mentioned actions,  we calculate  scalar field and  canonical normalize  potential . In  section 4, also by using  scalar field and  canonical normalize  potential we obtain the slow- roll parameters. Also, here we investigate the swampland condition. In section 5, we calculate the upper limit of $n$ and achieve  the $\alpha$, $\beta$ and $\gamma$ coefficients of the $f(R)$ theory. Also, we compare the corresponding obtained results with 2018 plank data. In last section, we discuss  the result of theory and also have some conclusion.
\section{Review of polynomial plus logarithmic correction of $f (R)$}
As we know the subject of dark energy may be origin of the late-time cosmic accelerated expansion. In that case, there are several successful model to describe dark energy. For example the dynamical dark energy model can be described by scalar field , modified matter (exotic fluid) and modified the general relativity as $f(R)$ or $F(R)$ models. Here, in order to investigate the inflation theory and arrange some parameter of theory from WGC point of view, first we assume the following action \cite{Sadeghi,Starobinsky2007},

\begin{equation}\label{5}
S=\int d^4x\sqrt{-g}( \frac{1}{2k^2}f(R)+L_{m} )
 \end{equation}

where $ k^{2}=8\Pi G=M_{pl} $ , $ M_{pl}$ is the reduced mass of Planck and $ L_{m}$ is the lagrangian density of matter. First of all we will try to explain the modified  $f (R)$ gravity which is play important role to describe dark energy and cosmic acceleration. It means that as a mentioned before one of the dynamical dark energy model are modification of general relativity which is form of $f (R)$ and it is a function of Ricci scalar.  So, generally one can say that the mentioned  modification is an alternative way for the dark energy as a cosmic acceleration.  Therefore,  different correction $ f (R)$  term is responsible for the quantum gravity theory. The correction term may be appear by  polynomial or logarithmic corrected term. Here, we mentioned that the ordinary function of $ f (R)$ without polynomial and logarithmic terms will be useful for the investigation of netron stars with strong magnetic field. The logarithmic correction may be useful for the effect of gluon in non- flat space - time and some cosmological model \cite{Sadeghi,Alavirad}. So, here  we try to consider more general form of  $ f (R)$  with polynomial plus logarithmic terms  which is given by,
\begin{equation}\label{6}
f(R)=R+\alpha R^2+\beta R^n +\gamma R^2 ln\gamma R
 \end{equation}

 where $ \alpha$, $\beta$ and $ \gamma$ are arbitrary constant. The most important things here is to arrange $n$ in equation (6). In order to specify such $n$ in power we first need to calculate  a series of cosmological parameters. Also, in second  step we need to use the   swampland  and slow-roll conditions. Also, we study the upper bend and investigate different $ \alpha$, $\beta$ and $ \gamma$  and examine the corresponding $f(R)$ function with suitable $n$. In that case we show that the obtained results from $n$ agree by results of literature. About  different values of $ \alpha$, $\beta$ and $ \gamma$ have already been mentioned in many articles \cite{Huang,Saidov}. Meanwhile, we want generally to examine our own function $f(R)$ and show what is the exactly $n$. Different values for the corresponding parameters in $f(R)$ theory can be compare to the obtained results from different papers  which are worked by researcher. In that case we take such exact values of parameters and obtain the cosmological quantities, which is compared by some excremental data. For this reason in next section we will try to investigate the scalar field and arrange the suitable form of potential.

\section{The corresponding  scalar field and potential in  $ f (R)$ modified gravity}

First of all we are going to use the general form of action which is given by,

\begin{equation}\label{7}
S=\frac{1}{2K^2}\int d^4x\sqrt{-g}f(R)
 \end{equation}
As you know $ f(R)= R + F(R)$, if $ F,_{RR}\neq 0 $, the corresponding action will be as,
\begin{equation}\label{8}
S=\int d^4x\sqrt{-g}( \frac{1}{2k^2}\varphi R - U(\varphi) )
 \end{equation}
 We note here the potential play an important role in weak gravity conjecture. Also, we know that  WGC cover two interesting condition as swampland and landscape. In order to investigate one of them as  swampland condition. we arrange the corresponding  potential in $f(R)$ model. So, we first move from Jordan to Einstein frame,  in this case we need some transformation which is known  conformal transformation\cite{Nojiri,Linder}. So by using the conformal transformations of metric\cite{Magnano}, we have following equation,
\begin{equation}\label{9}
\widetilde{g}_{\mu\nu}=f'(R)g_{\mu\nu}=((1+(2\alpha+\gamma)R+2\gamma R ln \gamma R)+n\beta R^{n-1})g_{\mu\nu}
 \end{equation}
So in that case the corresponding action will be as,
\begin{equation}\label{10}
S=\int d^{4}x\sqrt{-g}(\frac{1}{2k^2}\widetilde{R}-\frac{1}{2}\widetilde{g}^{\mu\nu}\nabla_{\mu}\phi \nabla_{\nu}\phi - V(\phi))
 \end{equation}
which is called Einstein Hilbert action. In order to calculate the canonical potential, we expand the logarithmic term and choose the largest term. So, by solving the following equation,
\begin{equation}\label{11}
\varphi=1+ \frac{\partial F(\chi)}{\partial \chi}
 \end{equation}
we have,
\begin{equation}\label{12}
\chi=(\frac{n}{n+2})^{(\frac{1}{n+1})}\frac{1}{\gamma}\varphi^{(\frac{1}{n+1})}
 \end{equation}
 where
\begin{equation}\label{13}
\varphi=\exp{\sqrt{\frac{2}{3}}\phi}
 \end{equation}
Now, we use general form of $f(R)$ as $F(\phi)$ and arrange the corresponding potential as,
\begin{equation}\label{14}
V(\phi)=\frac{(\varphi-1)R(\phi)- F(R(\phi))}{2\varphi^{2}}
 \end{equation}
The above equation with modified $f(R)$ gravity help us to obtain the potential as,
  \begin{equation}\label{15}
V(\phi)\simeq - \frac{1}{2\gamma K^2 }(\frac{n}{n+1})^{\frac{2}{n+1}}\frac{1}{n+1}(ln(\frac{n}{n+2})+\sqrt{\frac{2}{3}}\phi K)(\exp({-\sqrt{\frac{2}{3}}\frac{2n}{n+1}\phi}K))
 \end{equation}
So, the following section we take the above potential and   apply the swampland conjecture. In that case we achieve some information about $r$ and $n_s$ with their  figures. These help us to arrange the power of $R$ and also specify $\alpha$ , $\beta$ and $\gamma$.

 \section {The swampland conjecture}
In order to investigate the swampland condition,  we need to consider the equation (3). In that case,  we take equation (15) and rewrite the derivative  potential as,

\begin{equation}\label{16}
 \begin{split}
V'(\phi)=&-(\frac{(4\sqrt{6}+n^2(\sqrt{6}(9+n)-4 n^2 (4+n)K\phi+4n(5\sqrt{6}-4K\phi))}{6\gamma(1+n)^2 (2+n)^2 K})\\
&((\frac{n}{1+n})^{\frac{2}{2+n}}(\exp{-\frac{2n\phi K \sqrt{\frac{2}{3}} }{1+n}}))
 \end{split}
 \end{equation}
So, the equation of (3) and above potential lead us to obtain following relation,
  \begin{equation}\label{17}
\frac{K(-\sqrt{6}(4+n(4+n)(5+n)+4n(2+n)^{2}K\phi)}{(1+n)(-6(3+n)+\sqrt{6}(2+n)^{2}K\phi)} \geq c
 \end{equation}
 Now, by using the first swampland condition, we obtain  known cosmological parameters as scalar spectral index $n_{s}$ and  tensor to scalar ratio $r$ \cite{Starobinsky2007}.  First, we have to write two following equations which are correspond to  $n_{s}$ and $r$,

  \begin{equation}\label{18}
n_{s}=1-6\varepsilon+2\eta
 \end{equation}

  \begin{equation}\label{19}
r=16\varepsilon
 \end{equation}
 Here, we see that to arrange the above two corresponding parameters, we need some information from slow-roll and  swampland conditions. So, the first and second slow-roll conditions are given by \cite{Huang,Sadeghi,Garg},

 \begin{equation}\label{20}
\varepsilon=\frac{M_{pl}^{2}}{2}(\frac{V'}{V})^{2}
 \end{equation}
and
  \begin{equation}\label{21}
\eta=M_{pl}^{2}(\frac{V^{\prime\prime}}{V})
 \end{equation}
The obtained potential $(15)$ and slow-roll parameters from $(20)$ and $(21)$, one can obtain $\varepsilon$ and $\eta$ as,

 \begin{equation}\label{22}
\varepsilon=\frac{1}{2}(\frac{\sqrt{6}(4+n(4+n)(5+n))-4n(2+n)^{2}K\phi)^{2}}{(1+n)^{2}(-6(3+n)+\sqrt{6}(2+n)^{2}K\phi)^{2}})
 \end{equation}
and
 \begin{equation}\label{23}
\eta=8n(\frac{-3(4+n(14+n(7+n)))+\sqrt{6}n(2+n)^{2}K\phi)}{3(1+n)^{2}(-6(3+n)+\sqrt{6}(2+n)^{2}K\phi)}
 \end{equation}
 We note here, the swampland condition and slow-roll parameters lead us to obtain the scalar spectral index $n_{s}$ and the tensor to scalar ratio $r$,
\vspace*{0.5cm}
\begin{equation}\label{24}
 \begin{split}
n_{s}=&1-6\frac{1}{2}(\frac{\sqrt{6}(4+n(4+n)(5+n))-4n(2+n)^{2}K\phi)^{2}}{(1+n)^{2}(-6(3+n)+\sqrt{6}(2+n)^{2}K\phi)^{2}})
+\\&2*8n(\frac{-3(4+n(14+n(7+n)))+\sqrt{6}n(2+n)^{2}K\phi)}{3(1+n)^{2}(-6(3+n)+\sqrt{6}(2+n)^{2}K\phi)}
 \end{split}
 \end{equation}

\begin{equation}\label{25}
r=16(\frac{1}{2}(\frac{\sqrt{6}(4+n(4+n)(5+n))-4n(2+n)^{2}K\phi)^{2}}{(1+n)^{2}(-6(3+n)+\sqrt{6}(2+n)^{2}K\phi)^{2}}))
\end{equation}

\vspace*{1.5cm}
Here, first of all we use  two equations (24) and (25)  which are functions of $\phi$ and $n$.  In the second step, we obtain the corresponding field $\phi$ in terms of parameters $ n $ and $ n_{s}$ in equation (24),  and also here, we achieve the field $\phi$  in terms of $r$ and $n$ in equation (25).
Then by combining the above expressions  and using the first swampland condition with calculation  in equation (17),  one can obtain the values $ c $. So,  we note here, the values  $ c $ from  the first swampland condition can be written by scalar spectrum index $n_s$ and the tensor to scalar ratio $r$. Now we take advantage from Plank 2018 data for $n_s$ and $r$. The swampland condition lead us to put such values in two  mentioned above equation and obtain the corresponding $n$ in $f(R)$ gravity. Also, here we note that the above calculation help us to arrange two values of $n$ as 0.1541 and 0.0034 which are upper and lower bound respectively. Now we back to investigate the above obtained results of cosmological parameters as $n_{s}$ and $r$ from several figures.\\
 In that case, we draw different values of $c$ in terms of  $n_{s}$ , where $K$ and $n$ are flexed and coming from 2018 Plank data and above calculation respectively. Generally, we want to know  how the swampland conditions  guarantee the cosmological parameters and also other parameters of theory will be   matched by Plank 2018 data.

 \begin{figure}[h!]
 \begin{center}
 \subfigure[]{
 \includegraphics[height=9.5cm,width=8cm]{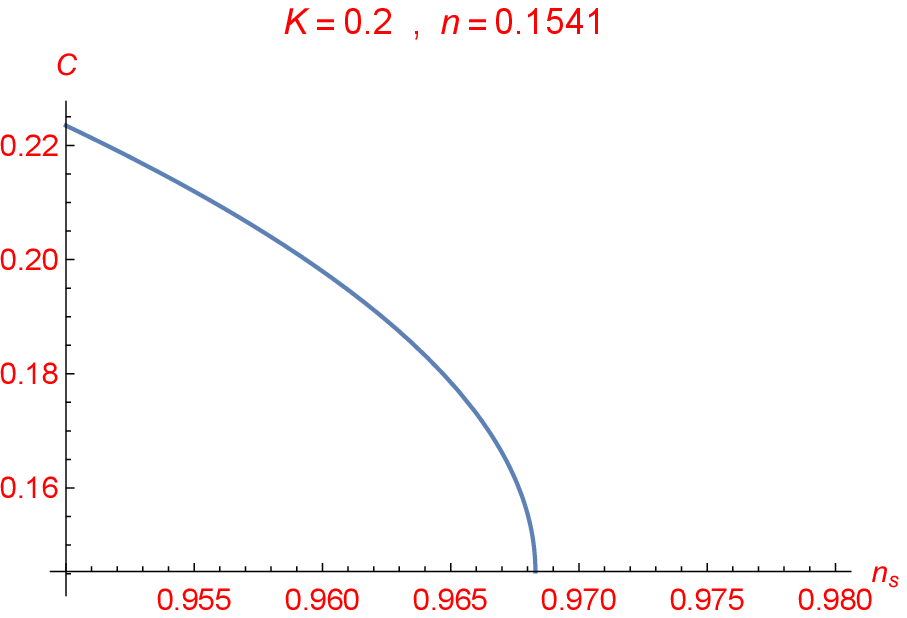}
 \label{1a}}
 \subfigure[]{
 \includegraphics[height=9.5cm,width=8cm]{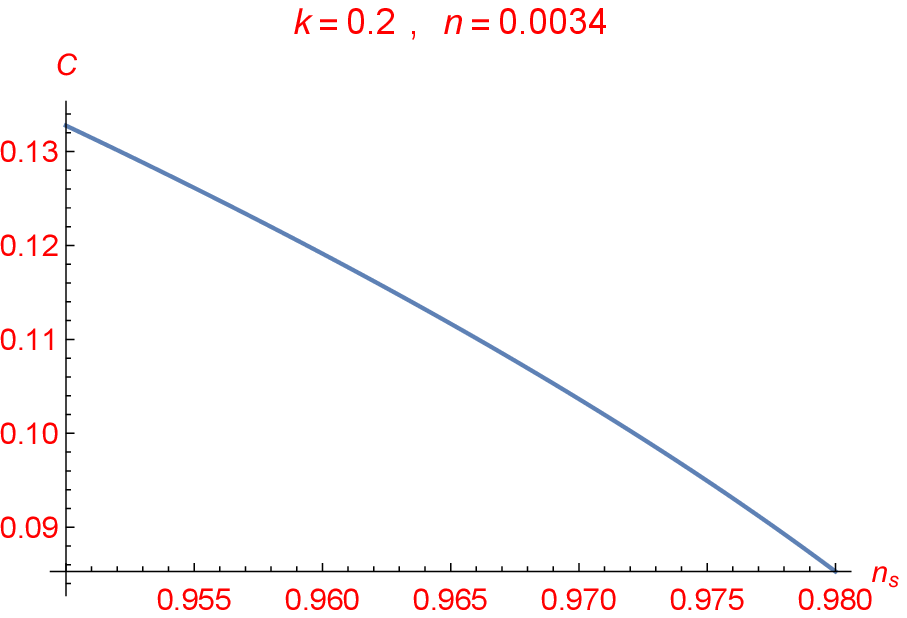}
 \label{1b}}
 \caption{\small{We  plotted different values of $c$ in terms of  the parameter $n_{s}$ with upper and lower bound of $n$, K also is fixed and given by Planck 2018, as shown in the figure $1a$ and $1b$}}
 \label{1}
 \end{center}
 \end{figure}
\newpage
Also, here we use equation (25) and second swampland condition (4), we obtain the values of $c'$ in terms of the tensor to scalar ratio $r$.
Now,  we plot graph $ c'$ in terms of $ r $, we will see how much this function changes.
\begin{figure}[h!]
 \begin{center}
 \subfigure[]{
 \includegraphics[height=9.5cm,width=8cm]{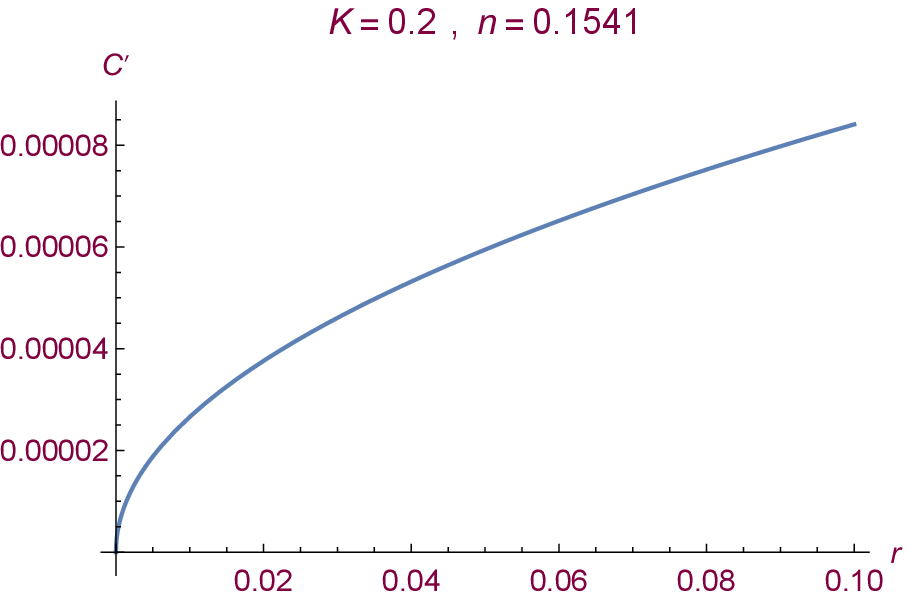}
 \label{2a}}
 \subfigure[]{
 \includegraphics[height=9.5cm,width=8cm]{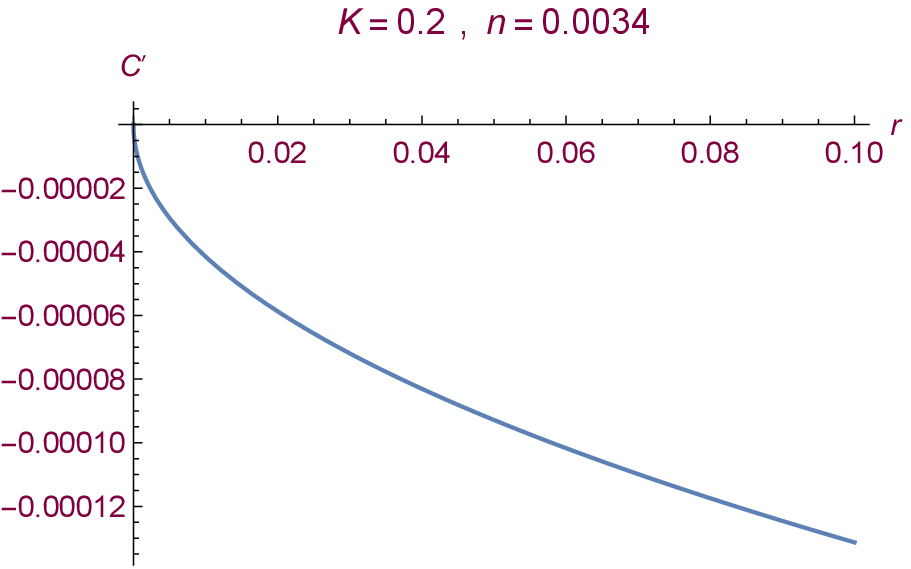}
 \label{2b}}
  \caption{\small{The variation of  $c'$ in terms of  parameter $r$ ,  K is given by Planck 2018 and the upper and lower bound of $n$ are obtained by calculation}}
 \label{2}
 \end{center}
 \end{figure}

As we mentioned before we wrote tensor to scalar ratio $r$ in terms of field in equation (25) and combined the corresponding equation with second swampland condition (4). So,   we obtained the values of $c'$ in terms of the tensor to scalar ratio $r$, see figures $2a$ and $2b$.
The figures $2a$ and $2b$ shown that the  permissible values are as stated in Planck 2018.

Also next step, we rewrite tensor to scalar ratio $r$ in terms of the scalar field  and combine  this relation with the first swampland condition (17). So we obtain the values of $c$ in terms of the tensor to scalar ratio.
Now, if we plot graph $ c$ in terms of $ r$, we can see how much this function changes.
\begin{figure}[h!]
 \begin{center}
 \subfigure[]{
 \includegraphics[height=9cm,width=8cm]{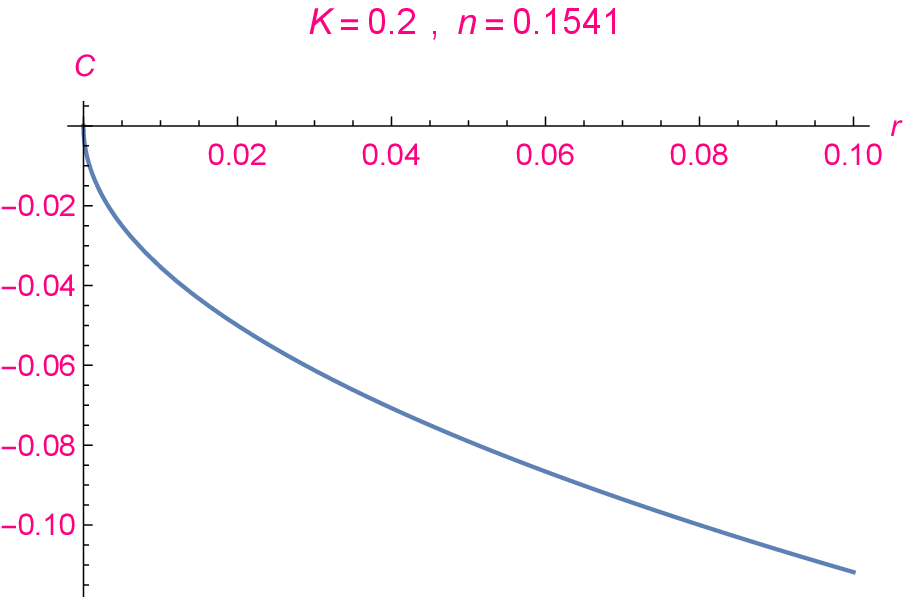}
 \label{3a}}
 \subfigure[]{
 \includegraphics[height=9cm,width=8cm]{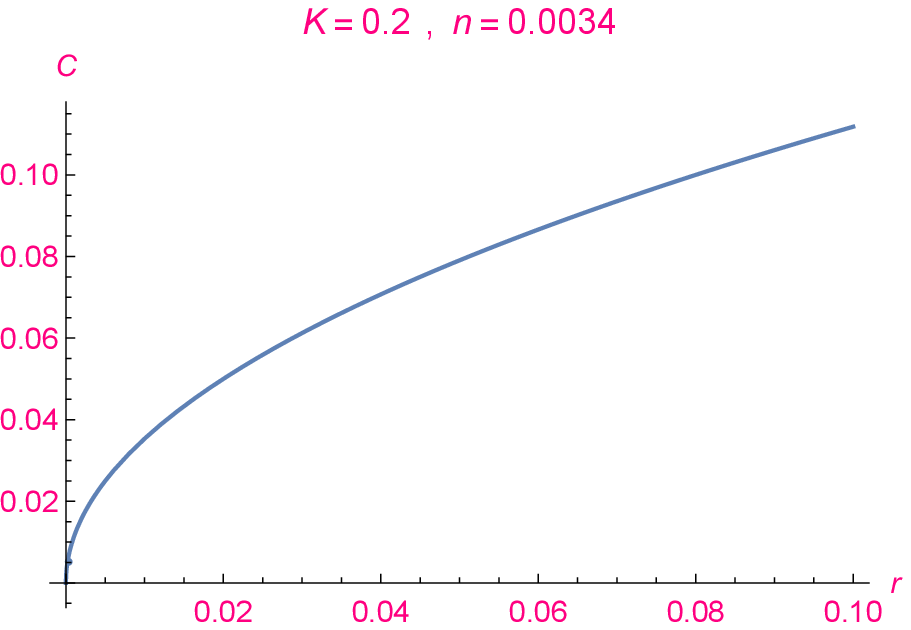}
 \label{3b}}
 \caption{\small{The variation of  $c$ in terms  $r$ with K and $n$ (upper and lower bound) are  given by Planck 2018 and  calculation respectively}}
 \label{3}
 \end{center}
 \end{figure}\\\\\\

As we know in this section and previous section  different swampland conditions give us different   diagrams .  In the following we will try to use
 relation between  cosmological parameters as discussed before, namely the scalar spectral index and the tensor to scalar ratio. So, in order to show that our result match to plank 2018 data, next step one can draw the variation $n_{s}$ in terms of $r$.
 \begin{figure}[h!]
 \begin{center}
 \subfigure[]{
 \includegraphics[height=8cm,width=8cm]{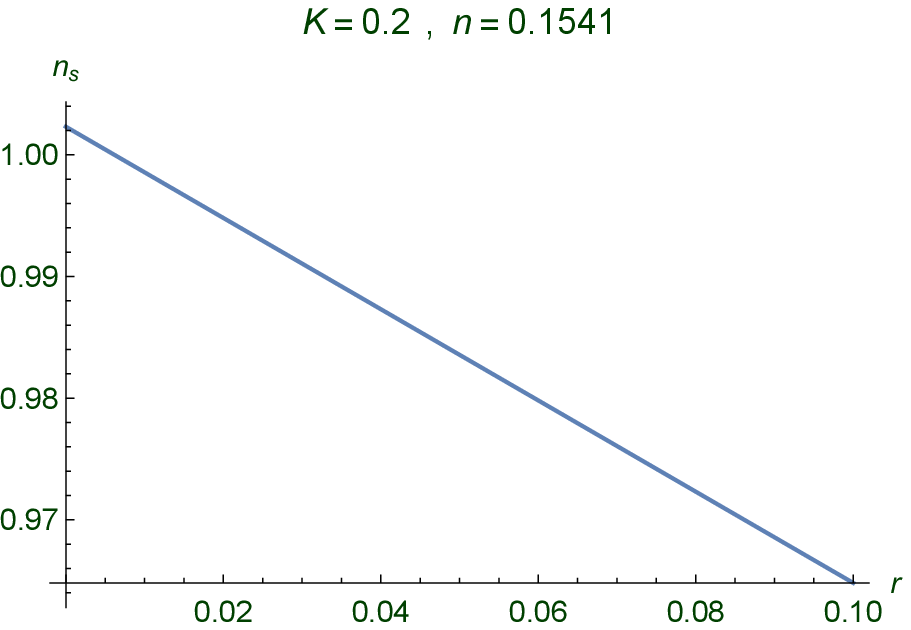}
 \label{4a}}
 \subfigure[]{
 \includegraphics[height=8cm,width=8cm]{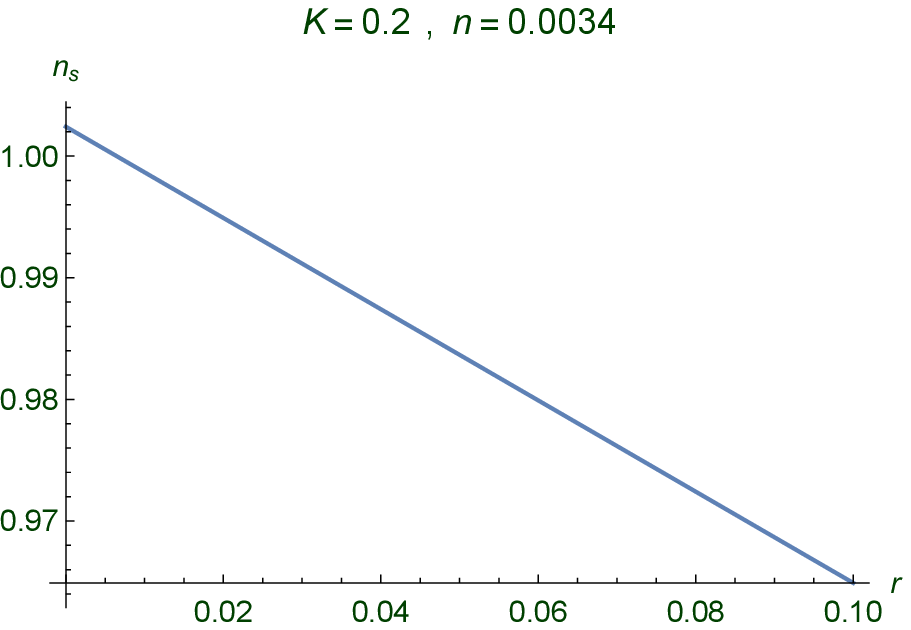}
 \label{4b}}
  \caption{\small{We have plotted the variation of $n_{s}$  in terms of  $r$. }}
 \label{4}
 \end{center}
 \end{figure}
\newpage{}

Here,  by using the equations (3), (4) and (24)  we obtain another constraint on the inflaton field. In that case we calculate $c^{2}c'^{2}$ and see how the $c$ change according to different values of $c'$.
\begin{figure}[h!]
 \begin{center}
 \subfigure[]{
 \includegraphics[height=8cm,width=8cm]{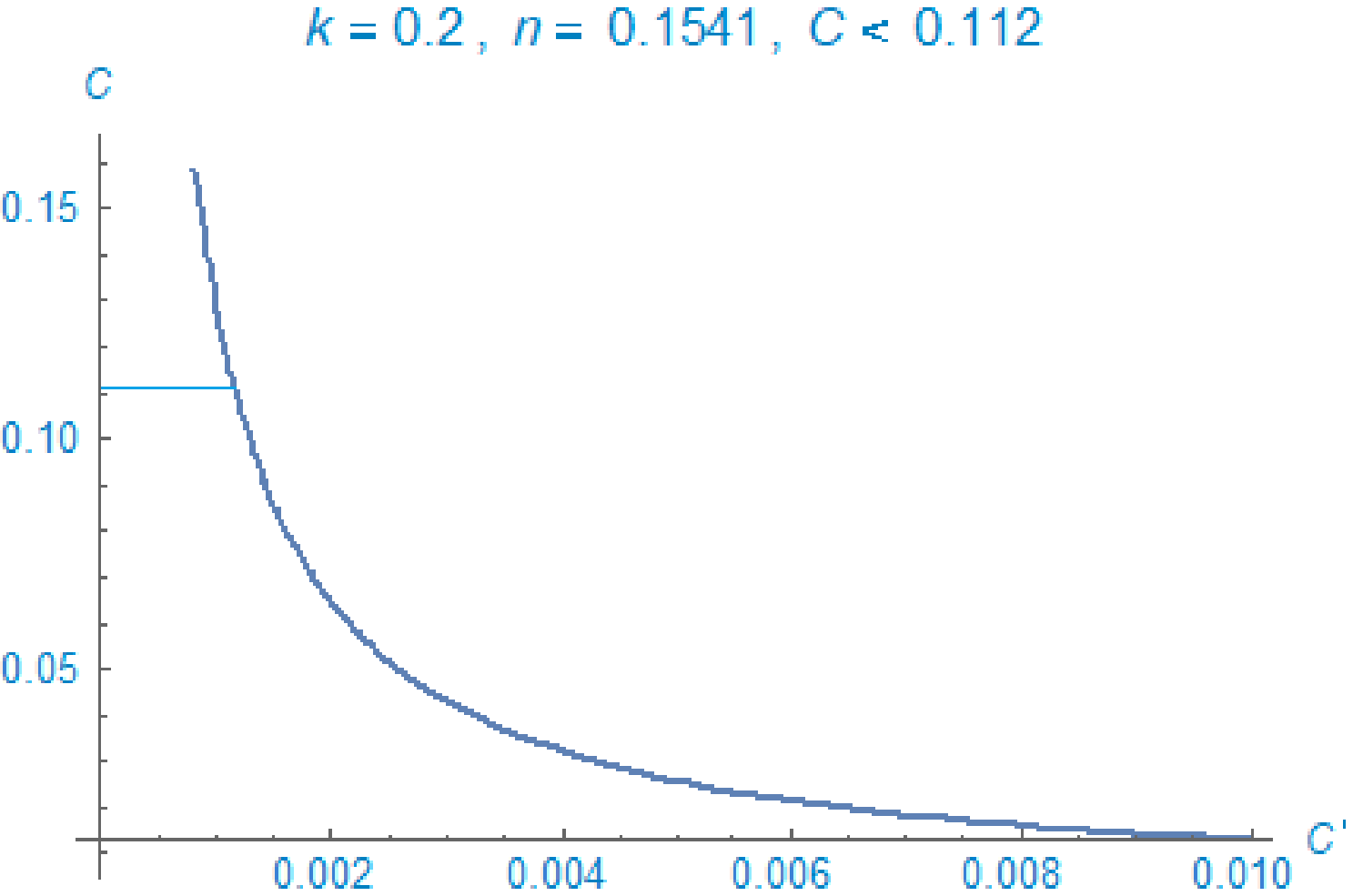}
 \label{5a}}
 \subfigure[]{
 \includegraphics[height=8cm,width=8cm]{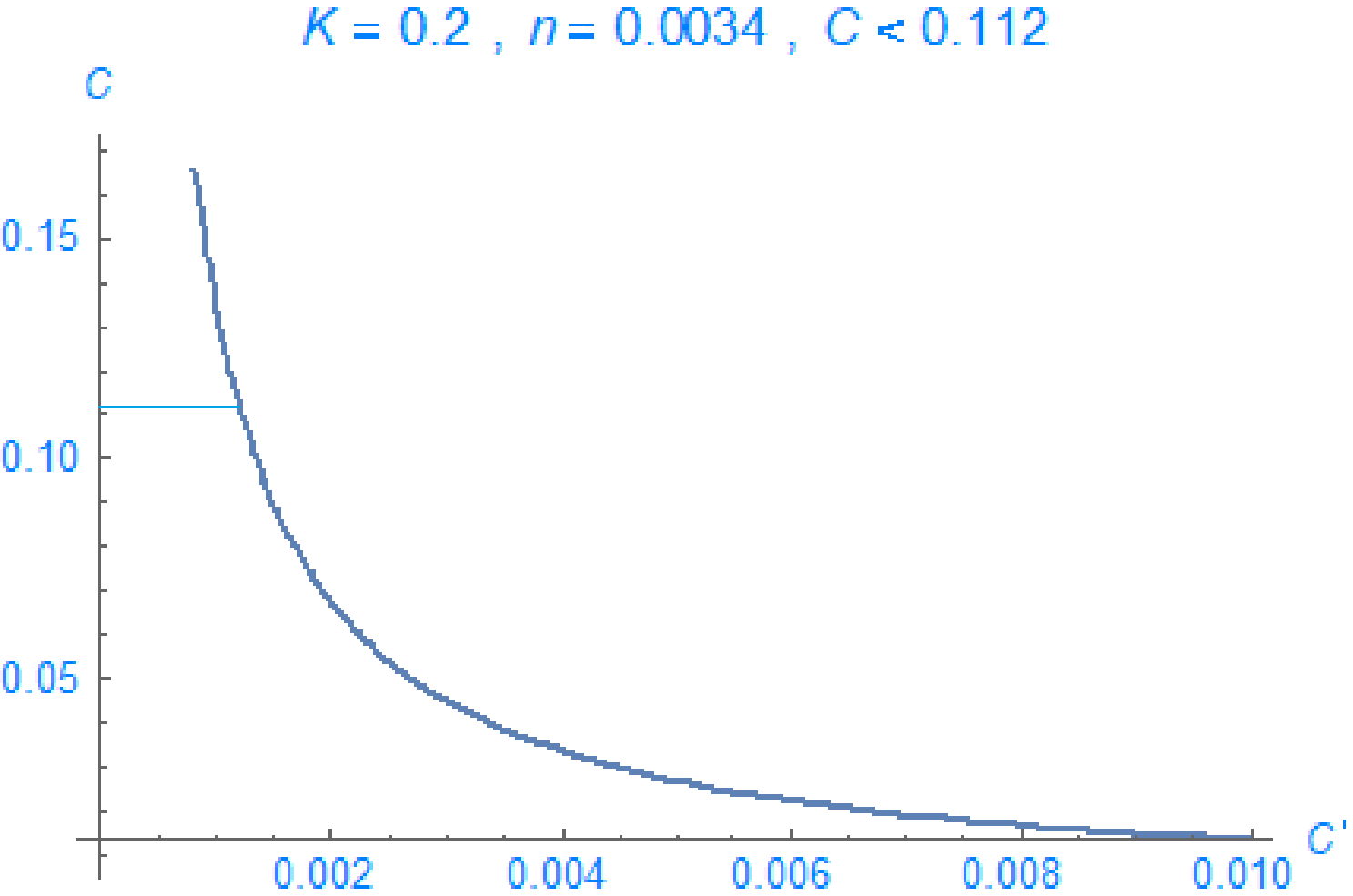}
 \label{5b}}
  \caption{\small{We have plotted the values of $c$  with respect to $c'$ with  given variables $ n$.}}
 \label{5}
 \end{center}
 \end{figure}
\newpage{}

The $K$ is given Plank 2018, and $n$ known upper and lower bound. In that case our permissible values are as stated in Planck 2018, as shown in the figure $5a$ and $5b$
according to the  condition  $f(n)\gg (c^{2})(c'^{2})$ with corresponding $n$ the permissible limit for the swampland condition is  $ c \ll 0,112$. Two corresponding figure as $5a$ and $5b$ complectly are satisfied by the condition of  $ c \ll 0,112$. Here, we mention that for the obtaining  the $ c\ll 0,112$,  we used before equations $(24)$, $(25)$ and $(17)$. So genially one can say that the  results from calculation and figures cover each other. By using mentioned  values  $n$ and $n_{s}$ , the permissible range  $c$ and  $c'$ are  specified by the corresponding figures.

\section{ The calculation of upper bound $ n$,  $\alpha$, $\beta$ and $\gamma$ coefficients theory}
In order to study n, $\alpha$, $\beta$ and $\gamma$ coefficients, we use the swampland distance conjecture equation $(1)$, $(17)$ and $(24)$. On the other hand, the value of $|\Delta \varphi|\sim1$ for different values of $n$ and $n_{s}=0.9649$, the first swampland condition must be satisfied by following relation.
\begin{equation}\label{26}
\Delta u \simeq u
 \end{equation}
Now, we use the above equation about scalar spectral index and tensor to scalar ratio. In that case, we employ the value of $n_{s}$ from plank 2018 data and put the equation $(17)$, and obtain the upper bound of $n$. We achieved the different value with respect to the available equation for $n$. The most of results are satisfied by the above mentioned $n$ as $n=0.1541$. We note here, the upper and lower limit of value of $n$, the relation of $n_{s}$ and $r$ help to compute $\alpha$, $\beta$ and $\gamma$ coefficients of the corresponding theory. The coefficients are arranged with upper limit of $n$ and $n_{s}$ (from plank 2018) by the following table.
\begin{figure}[h!]
 \begin{center}
 \subfigure[]{
 \includegraphics[height=8cm,width=8cm]{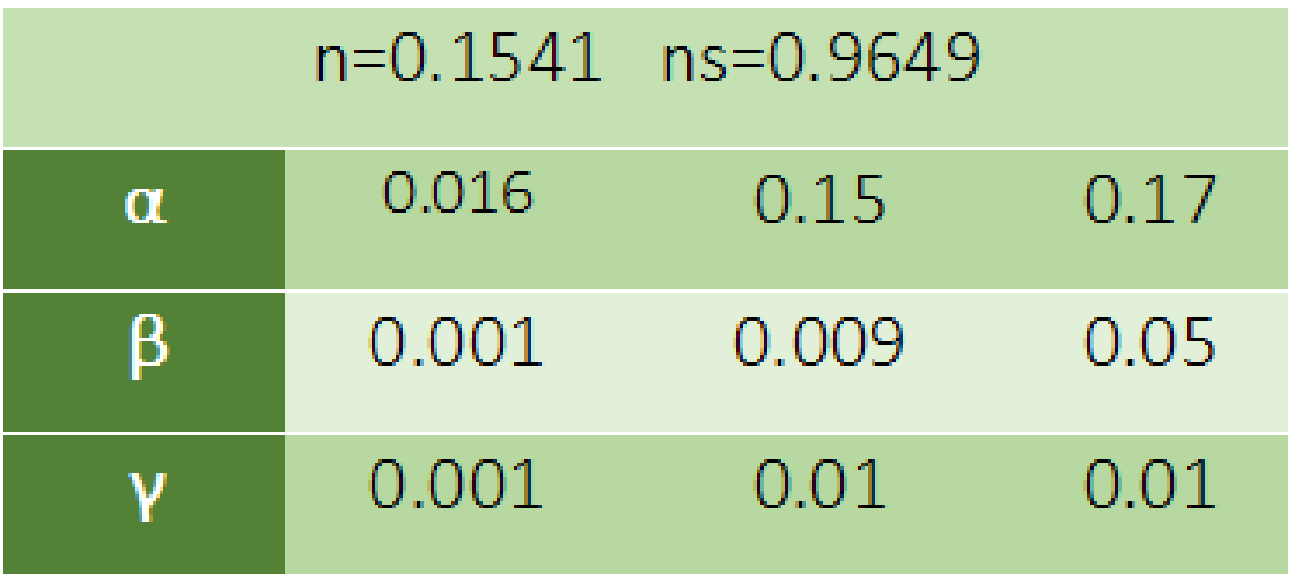}
 \label{6a}}
 \caption{\small{The values of the coefficients obtained for $n$ and $n_{s}$. (Different values have been calculated for these coefficients with respect to the two above parameters, we have selected only a few of these permissible values between zero and one).}}
 \label{6}
 \end{center}
 \end{figure}
So, the above calculation lead us to show different value of coefficients in table $6$. Generally, we obtain different coefficients and we only keep some values between zero and one. Because,  some physical arguments and obtained results from literature give us opportunity to keep such values\cite{Kehagias,Akrami}.  We note here, the first swampland condition satisfied by $ c \simeq 1$. But, here  the suitable coefficients $\alpha$, $\beta$ and $\gamma$ ,  $n$,  some  2018 plank data and information from mentioned figures about  $f(R)$ theory lead us to obtain approximately values for $c$ as $0.11$ (upper bound of $n$). This corresponding $c$ are completely satisfied by Ref.s.\cite{Kehagias,Akrami}.
Finally, we check all values obtained for $n$, $n_{s}$ and $r$ in figure(7). In fig (7a),  we take $n=0.1541$ and see the variation of $n_{s}$  and $ r $ with respect to corresponding $n$.

\begin{figure}[h!]
 \begin{center}
 \subfigure[]{
 \includegraphics[height=8cm,width=8cm]{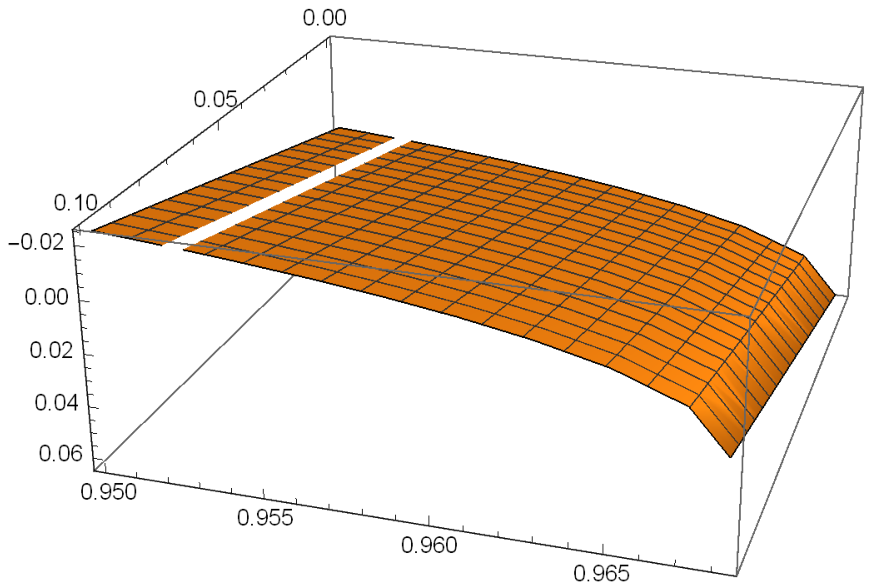}
 \label{7a}}
 \subfigure[]{
 \includegraphics[height=8cm,width=8cm]{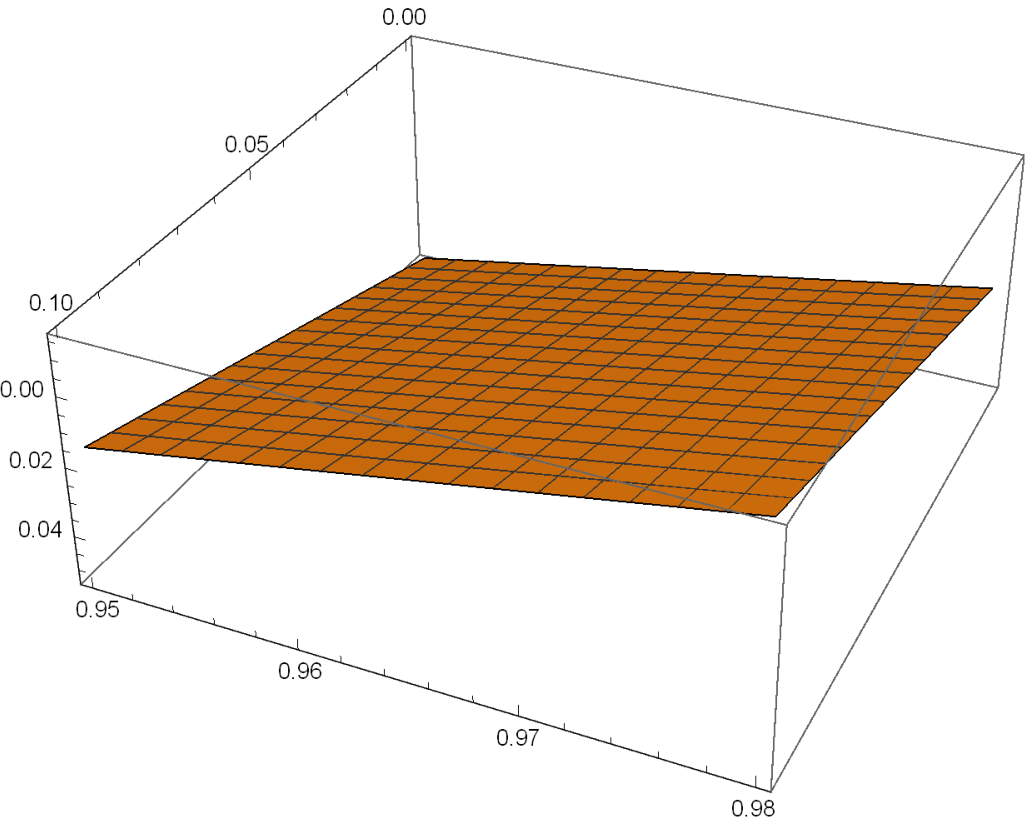}
 \label{7d}}
\caption{\small{We have plotted the graph above using the permissible values of the parameters $n$, $r$ and $n_{s}$ expressed by the relationships. Which shows well the range of our guess.$7a: n=0.1541$ and $7b : n=0.0034$}}
 \label{7}
 \end{center}
 \end{figure}

\newpage{}
\section{Conclusions}
Recently specific conditions such as the swampland condition are used to investigate the  low-energy theories. So, we took advantage from that conjecture  to study $f(R)$ theory. So, first we introduced  inflation model with its own function, which is a polynomial plus a logarithmic terms. In that case we  used   Einstein Hilbert's action and  calculated the scalar field and the canonical normalization potential. We also here  examined a number of cosmological parameters and investigated the validity $f(R)$ theory. By using scalar spectral index, tensor to scalar ratios and slow roll relations we also examined the validity of the swampland condition. And we show that our inflation model is consistent with the swampland conjecture. In that case, we  used equations (24) and (25) with respect to 2018  planck data and obtained  the upper and lower bound of $n$ as \textcolor{red}{0.15} and \textcolor{red}{0.0033}. We adjusted our computational values to Planck's values of 2018 and finally calculated each of the coefficients  $\alpha$, $\beta$ and $\gamma$ with respect to the upper limit $ n$ for  the corresponding model. It may be interesting to investigate the swampland conjecture to the corresponding model with constant-roll instead of slow-roll conation.

\end{document}